\documentstyle[12pt,a41,epsfig,cite,portland,rotating]{article}

\newcommand{\beq}{\begin{equation}}
\newcommand{\eeq}{\end{equation}}
\newcommand{\bea}{\begin{eqnarray}}
\newcommand{\eea}{\end{eqnarray}}

\newcommand\MeV{\,\mbox{MeV}}
\newcommand\GeV{\,\mbox{GeV}}

\newcounter{lin}

\begin{document}
\begin{titlepage}

\begin{flushleft}
DESY 07--227 \hfill {\tt arXiv:0802.0408 [hep-ph]} 
\\
SFB-CPP-08-12 \\
\end{flushleft}

\vspace{3cm}
\noindent
\begin{center}
{\LARGE\bf Higher Twist Contributions to the Structure} \\
\vspace*{2mm}
\noindent
{\LARGE \bf Functions \boldmath $F_2^{p}(x,Q^2)$
and $F_2^{d}(x,Q^2)$ at Large $x$}  \\
\vspace*{2mm}
\noindent
{\LARGE \bf \boldmath at Higher Orders}
\end{center}
\begin{center}

\vspace{2.0cm}
{\large Johannes Bl\"umlein and  Helmut B\"ottcher}

\vspace{1.5cm}
{\it 
Deutsches Elektronen Synchrotron, DESY}\\

\vspace{3mm}
{\it  Platanenallee 6, D--15738 Zeuthen, Germany}\\

\vspace{3cm}
\end{center}
\begin{abstract}
\noindent
The higher twist contributions to the deeply inelastic structure functions
$F_2^{p}(x,Q^2)$ and $F_2^{d}(x,Q^2)$ for larger values of the Bjorken variable $x$
are extracted extrapolating the {twist--2} contributions measured in the large $W^2$
region to the region $4 \GeV^2 \leq W^2 \leq 12.5 \GeV^2$ applying target mass 
corrections. We compare the results for the NLO, NNLO and N$^3$LO analyzes and include
also the large $x$ at N$^4$LO to the Wilson coefficients. A gradual lowering of the 
higher twist contributions going from NLO to N$^4$LO is observed, which stresses the 
importance of higher order corrections.
\end{abstract}

\end{titlepage}

\newpage
\sloppy


\vspace{1mm}\noindent
Deeply inelastic structure functions contain higher twist corrections \cite{HT} both
in the region of large and small values of the Bjorken variable $x$ 
\cite{HT1a,HT1,HT2,HT3}.
While the leading twist sector both for unpolarized and polarized deeply inelastic 
scattering is well explored within perturbative Quantum Chromodynamics (QCD) up to the
level of 3--loop, resp. 2--loop,  corrections \cite{THR1,THR2}, very little is known 
on the scaling violations
of dynamical next-to-leading twist correlation functions and the associated Wilson
coefficients \cite{HT}, even on the leading order level.~\footnote{In the small $x$ 
region saturation
corrections, see \cite{HT2,HT3}, are considered which also belong to the class of higher 
twist corrections, applying different formalisms. An important question to be addressed
here concerns sub-leading corrections, which may be large, cf.~\cite{BV}. This has been 
confirmed in a fixed order calculation to three loops in \cite{VERM1}. Recent investigations
of the small-$x$ behaviour of deeply inelastic structure functions account for these effects
\cite{ALTAR,CIAFA}.} 
In many experimental and 
phenomenological analyzes, cf.~\cite{HT1a,HT1}, higher twist contributions are parameterized 
by an `Ansatz' \cite{HT1a}, which is fitted accordingly. Within QCD this ad-hoc treatment  
usually cannot  be justified, performing at the same time a higher order analysis for the 
leading twist terms. Since neither the corresponding higher twist anomalous dimensions nor
Wilson coefficients were calculated, the data analysis has to be limited in the first place 
to the kinematic domain in which higher twist terms can be safely disregarded. 

In the case of flavor non-singlet combinations of structure functions this is widely the case 
in the region $Q^2 \geq 4 \GeV^2, W^2 \geq 12.5 \GeV^2$, as detailed analyzes of the large 
$W^2$ region show, cf. \cite{BBG}. In this region one may perform a three-loop QCD analysis,
which requires the $O(\alpha_s^2)$ Wilson coefficients \cite{ZN} and the 3--loop anomalous 
dimensions \cite{THR1}. The analysis can even be extended effectively to 4--loop order, since 
the dominant contribution there is implied by the 3--loop Wilson coefficient \cite{THR2}, 
parameterizing the yet unknown 4--loop anomalous dimension with a $\pm$ 100~\% error added to   
an estimate of this quantity formed as Pad\'e-approximation out of the lower order terms.
A comparison with the 2nd moment of the 4--loop anomalous dimension calculated in \cite{BC}
showed \cite{BBG} that the agreement is better than 20~\%, which underlines that the above 
approximation may be possible. We limit the QCD--analysis of the twist--2 contributions
to this representation since neither $\alpha_s(\mu^2)$ nor the splitting and coefficient functions 
are known beyond this level. 

The evolution equations are solved in Mellin-$N$ space. The non--singlet structure function at 
the starting scale of the evolution, $Q^2_0$, is given by
\begin{eqnarray}
\label{eq1}
F_2^{p,d; \rm NS}(N,Q^2) = \sum_{k=0}^\infty a_s^{k-1}(Q^2) C_{k-1}^{\rm NS}(N)
             f_2^{p,d; \rm NS}(N,Q^2)~,
\end{eqnarray}
with $C_k^{\rm NS}(N)$ the expansion terms of the non--singlet Wilson coefficient with 
$C_0(N) = 1$, $a_s(Q^2)= \alpha_s(Q^2)/(4\pi)$ and $f_2^{p,d; NS}(N,Q^2)$ the corresponding
combination of quark distributions, cf. \cite{BBG}. Here we identify both the renormalization and
factorization scale with $Q^2$. Beyond $O(a_s^3)$ dominant large $x$ contributions to the
Wilson coefficient were calculated in \cite{resum}.

The evolution equation for the quark densities to 4--loop order reads~:
\begin{eqnarray}
\label{eq2}
f_2^{p,d; \rm NS}(N,Q^2) &=& f_2^{p,d; \rm NS}(N,Q_0^2)
\left(\frac{a}{a_0}\right)^{-\hat{P}_0(N)/{\beta_0}}
\Biggl\{1 - \frac{1}{\beta_0} (a - a_0) \left[\hat{P}_1^+(N)
- \frac{\beta_1}{\beta_0} \hat{P}_0(N) \right] \nonumber\\ 
& & - \frac{1}{2 \beta_0}\left(a^2 - a_0^2\right) \left[\hat{P}_2^+(N) 
- \frac{\beta_1}{\beta_0} \hat{P}_1^+(N) + \left( \frac{\beta_1^2 - \beta_0 
\beta_2}{\beta_0^2}
\right) \hat{P}_0(N)   \right]
\nonumber\\ & &
+ \frac{1}{2 \beta_0^2} \left(a - a_0\right)^2 \left(\hat{P}_1^+(N)
- \frac{\beta_1}{\beta_0} \hat{P}_0(N) \right)^2
\nonumber\\ & &
- \frac{1}{3 \beta_0} \left(a^3 - a_0^3\right)
\Biggl[\hat{P}_3^+(N)
- \frac{\beta_1}{\beta_0} \hat{P}_2^+(N)  
+ \left(\frac{\beta_1^2 - \beta_0 \beta_2}{\beta_0^2}  
\right) \hat{P}_1^+(N)
\nonumber\\ & & 
+\left(\frac{\beta_1^3}{\beta_0^3} -2 \frac{\beta_1 \beta_2}{\beta_0^2} + 
\frac{\beta_3}{\beta_0} \right) \hat{P}_0(N) \Biggr] 
\frac{\left(a-a_0\right)\left(a_0^2 - a^2\right)}{2\beta_0^2}
\left(\hat{P}_1^+(N)-\frac{\beta_1}{\beta_0} \hat{P}_0(N) \right) 
\nonumber\\ & & \times
\left[\hat{P}_2(N) - \frac{\beta_1}{\beta_0} \hat{P}_1(N) - 
\left(\frac{\beta_1^2-\beta_0 \beta_2}{\beta_0^2} \right) 
\hat{P}_0(N) 
 \right]
\nonumber\\ & &
-\frac{\left(a-a_0\right)^3}{6 \beta_0^3} 
\left(\hat{P}_1^+(N)-\frac{\beta_1}{\beta_0} \hat{P}_0(N)  
\right)^3 \Bigg\}~.
\end{eqnarray}
Here $\hat{P}_k^+(N)$ denotes the $(k+1)$--loop anomalous dimension
and $\beta_k$ are the expansion coefficients of the QCD $\beta$-function, cf. 
\cite{BETA}, with 
\begin{eqnarray}
\label{eq_a}
\frac{d a(\mu^2)}{d \ln(\mu^2)} = - \sum_{k=0}^\infty \beta_k a^{k+2}(\mu^2)~.
\end{eqnarray}
Eq.~(\ref{eq_a}) is solved perturbatively to 4--loop order in the $\overline{\rm MS}$
scheme \cite{ALS} observing
the flavor matching conditions for the renormalization scale  $\mu$ at the thresholds 
$m_c = 1.5 \GeV$ and $m_b = 4.5 \GeV$, respectively,
to be able to compare to other measurements of $\alpha_s(M_Z^2)$, resp. $\Lambda_{\rm 
QCD}^{\rm N_f}$.  
To perform the data analysis the expression for the structure functions, (\ref{eq1}), is 
transformed back
to $x$--space by a numeric contour integral around the 
singularities of the problem in the complex $N$-plane.

In the analysis mass corrections have to be accounted for. These are the target 
mass  \cite{TARG} and heavy flavor corrections \cite{HEAV1,HEAV2,HEAV3}. While the former are significant, 
the latter ones amount only 1-2 \% at NLO and are expected to be even smaller in the yet 
unknown higher orders\footnote{First contributions which are relevant for the $O(\alpha_s^3)$ heavy flavor 
contributions 
to $F_2(x,Q^2)$ were calculated in \cite{CONST} for the region $Q^2  \gg m^2$. Under this kinematic condition 
the corresponding corrections to $F_L(x,Q^2)$ were calculated in \cite{FL}.}  
in the flavor non-singlet case. In the evolution equations (\ref{eq1},\ref{eq2}) the anomalous dimensions
and coefficient functions are represented in $N$-space \cite{ANCONT,HEAV2,NV,THR1,THR2}. Here we applied 
simplifications due to algebraic \cite{ALGEBRA} and structural relations \cite{STRUCT} between harmonic sums. 
Under the conditions 
mentioned above we perform the twist--2 analysis from leading order (LO) to 4--loop order (N$^3$LO) fitting the 
non-singlet $F_2^{p,d}(x,Q^2)$ world data, cf. \cite{BBG}, using {\tt MINUIT} \cite{MINUIT}.  We then extrapolate the 
results to the region $4 \GeV^2 \leq W^2 \leq 12.5 \GeV^2$ and determine effective  higher twist coefficients
$C_{\rm HT}(x,Q^2)$ given by
\begin{equation}
F_2^{\rm exp}(x,Q^2) = F_2^{\rm tw2}(x,Q^2) \cdot \left[
\frac{O_{\rm TMC}\left[F_2^{\rm tw2}(x,Q^2)\right]} 
                      {F_2^{\rm tw2}(x,Q^2)} 
+ \frac{C_{\rm HT}(x,Q^2)}{Q^2 [1 \GeV^2]}\right]~.
\end{equation}
Here $O_{\rm TMC}[~~]$ denotes the operator of target mass corrections.

QCD corrections beyond N$^3$LO are known in form of the dominant large-$x$ contributions 
to the QCD--Wilson coefficients \cite{resum}. Since these corrections do quantitatively 
only apply in the range of large $x$ we do not 
use them in the twist-2 QCD--fit, because here the data are mainly situated at lower 
values of $x$ where beyond 4--loop order 
other contributions to the Wilson coefficients, which are not calculated yet, are as 
important. Furthermore, the 4--loop anomalous dimensions were not calculated 
yet.
The leading large $x$ contributions are
given in terms of harmonic sums of the type $S_{1,1, \ldots, 1}(N)$ which obey a determinant representation \cite{BK}
in single harmonic sums $S_l(N)$, i.e. they are polynomials of single harmonic sums.\footnote{Related representations 
were given in \cite{SSU}.} One may calculate these sums recursively, \cite{BK}. The $n$-fold sum reduces to
the $(n-k)$-fold sums by 
\begin{equation}
S_{1, \ldots, 1}(N) = \frac{1}{n} \left[S_n(N) + S_{1}(N) S_{n-1}(N) + S_{1,1}(N) S_{n-2}(N) + \ldots \right]~. 
\end{equation}
To obtain the large $x$ behaviour we retain the terms
\begin{eqnarray}
S_{1}(N) &\propto& \ln(N) + \gamma_E \\
S_{l}(N) &\propto& \zeta_l,~~~ l \geq 2~,
\end{eqnarray}
as $|N| \rightarrow \infty$,
with $\gamma_E$ the Euler--Mascheroni number, and $\zeta_l$ 
the Riemann $\zeta$-function at integer values. We agree with the numerical 
parameterization given in Table~1, [10a].
We take into account the N$^4$LO terms in this approximation, which are added to the 
twist--2
fit results in N$^3$LO extrapolating to the region of lower values of $W^2$.

The effective higher twist distribution functions $C_{\rm HT}^{p,d}(x)$ extracted are shown in Figures~1 and 2 from
NLO to N$^4$LO.
Here we averaged over the values in $Q^2$ within the $x$--bins.~\footnote{We took the 
opportunity to re-bin
the data at very large $x$, if compared to \cite{BBG}, to optimize w.r.t. experimental errors.}
The leading twist terms are those given in \cite{BBG}, with the values of $\Lambda_{\rm QCD}^{(4)} = 265 \pm 27,
226 \pm 25, 234 \pm 26 \MeV$, resp. in NLO, NNLO, and N$^3$LO.
Both for the proton and deuteron data $C_{\rm HT}(x)$ grows towards large values of $x$, and takes values $\sim 1$ around
$x = 0.6$. The inclusion of higher order corrections reduces $C_{\rm HT}(x)$ to lower values with a gradually
smaller difference order by order. Yet for the highest bins, $x \geq 0.8$, the effect of the large $x$ resummation
terms is important. Earlier higher twist analyzes \cite{HT1} limited to the next-to-leading order corrections 
are thus corrected by factors of 2 and larger at large $x$ to lower values. In the present analysis we limited the 
investigation to the inclusion of the large $x$ terms in N$^4$LO which are still in the 
vicinity of a nearly
complete QCD analysis as outlined above. The present description is likely to be final 
for values of $x \leq 0.8$.
Beyond this range there are only few data. More data in this interesting region would be 
welcome and can be obtained
at planned high-luminosity colliders such as EIC \cite{EIC}. Data from measurements at JLAB cannot be included into the 
present analysis, since the kinematic region covered currently averages over the resonances applying duality, which
may lead to different and probably lower higher twist contributions, cf. \cite{HTJ}.  

In the present analysis we extracted the large $x$ dynamical higher twist contributions to the structure functions 
in a  model-independent way. It would be interesting to compare moments of the term 
$C_{\rm HT}(x)$ to lattice
results, which allow to simulate the moments of the corresponding higher twist correlation functions, in the future.

\vspace{3mm}
\noindent
{\bf Acknowledgment.}\\
This work was supported in part by by DFG Sonderforschungsbereich Transregio 9, 
Computergest\"utzte Theoretische Physik. For discussions we would like to 
thank S. Moch, V. Ravindran, and A. Vogt.



\newpage
\begin{figure}[t]
\begin{center}
\includegraphics[angle=0, width=14.0cm]{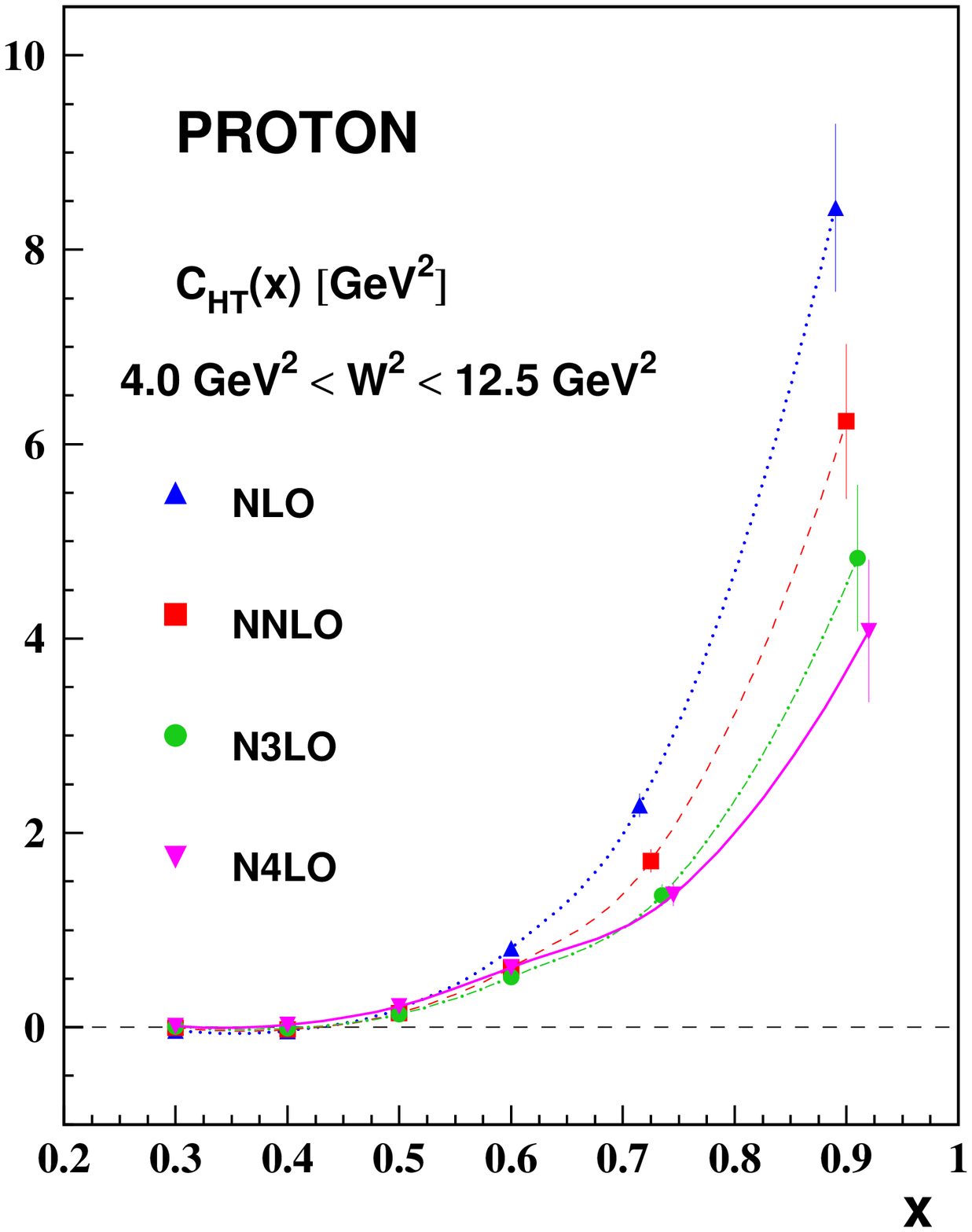} 
\end{center}
{\sf
\caption{\label{fig:HT_p}
Comparison of the higher twist coefficient $C_{\rm HT}(x)$ in the large $x$ region 
for the proton data as function of $x$ in a NLO (dotted line), NNLO (dashes line), 
N$^3$LO analysis (dash-dotted line) and adding the large $x$ terms in $O(\alpha_s^4)$ 
for the non-singlet QCD Wilson coefficient (full line). Some bin centers are slightly shifted for 
better 
visibility.}}
\end{figure}
\normalsize

\newpage
\begin{figure}[t]
\begin{center}
\includegraphics[angle=0, width=14.0cm]{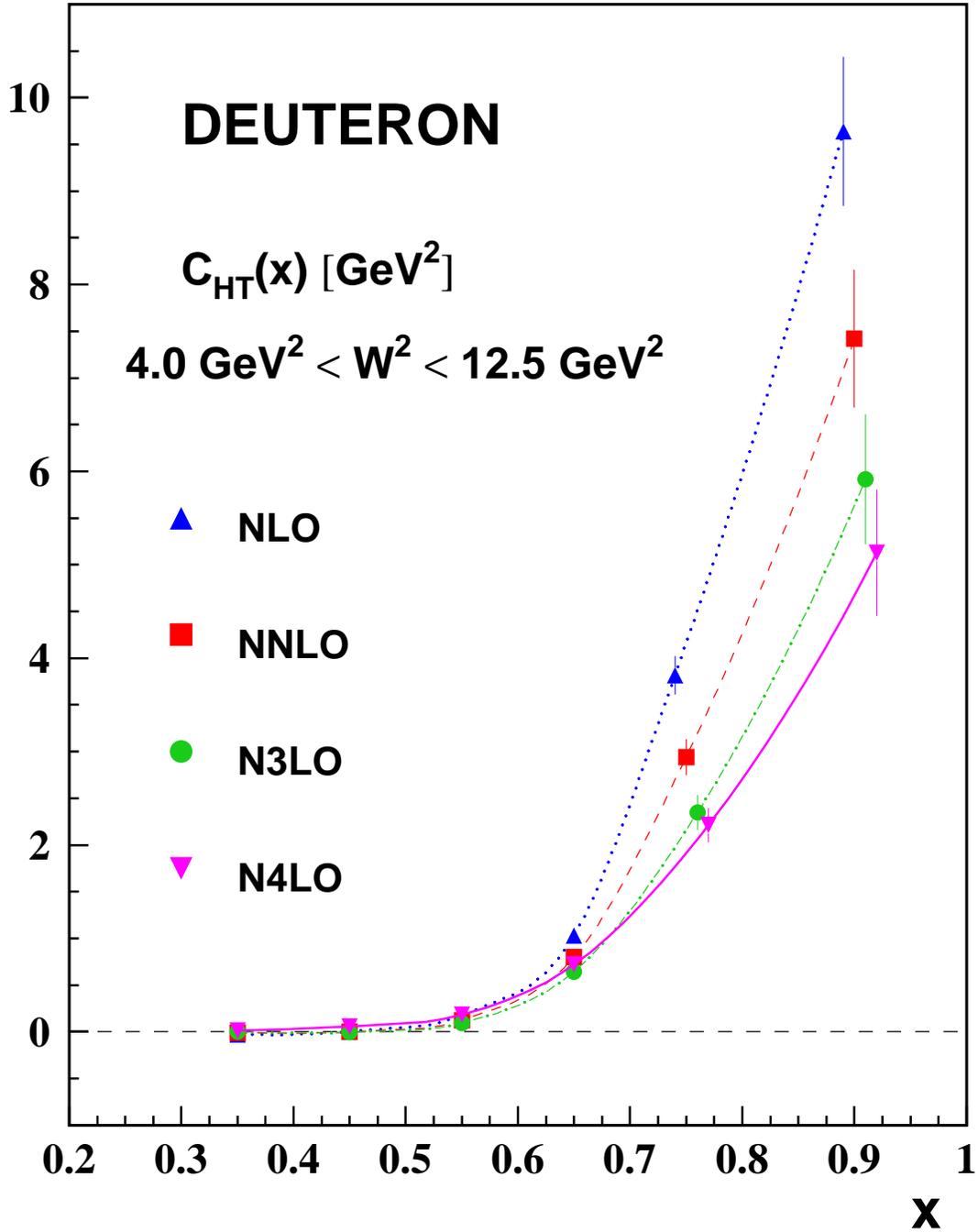} 
\end{center}
{\sf
\caption{\label{fig:HT_d}
The coefficient $C_{\rm HT}(x)$ for the deuteron data. The curves have the same meaning as
in Figure~1.}}
\end{figure}
\normalsize

\end{document}